\documentstyle[12pt]{article}
\begin{document}
\bibliographystyle{unsrt}
\newcommand{\bra}[1]{\left < \halfthin #1 \right |\halfthin}
\newcommand{\ket}[1]{\left | \halfthin #1 \halfthin \right >}
\newcommand{\be}{\begin{equation}}
\newcommand{\ee}{\end{equation}}
\newcommand{\vsig}{\mbox {\boldmath $\sigma$\unboldmath}}
\newcommand{\vep}{\mbox {\boldmath $\epsilon$\unboldmath}}
\newcommand{\fn}{\frac 1{E^i+M_N}}
\newcommand{\fs}{\frac 1{E^f+M_N}}

\title{\bf  The $\eta$ Photoproduction of Nucleons and The Structure of
The Resonance $S_{11}(1535)$ in the Quark Model}

\author{Zhenping Li
\\
Physics Department, Carnegie-Mellon University \\
Pittsburgh, PA. 15213-3890 }
\maketitle

\begin{abstract}
In this paper, we present our study on the $\eta$ photoproduction
based on the chiral quark model.  We find that quark model
provides a very good description of the $\eta$ production with much less
parameters,   and the threshold region is not a reliable
source to determine the $\eta NN$ coupling constant due to its strong
dependence on the properties of the resonance $S_{11}(1535)$.
We suggest that the systematic data in $E_{lab}=1.2 \sim 1.4$ GeV region
may help us to determine the $\eta NN$ coupling constant more precisely.
 The structure of the resonance $S_{11}(1535)$ is discussed,  we find that
the recent data from Mainz group bring the helicity amplitude much
closer to the quark model prediction.  However, more studies need to be
done to understand the large $\eta N$ branching ratio of the resonance
$S_{11}(1535)$.  Our results show that the quark model is a very good
approach to study the underlying structure of baryon resonances from the
meson photoproduction data.
\end{abstract}

\newpage
\subsection*{\bf 1. Introduction}

In our previous publication\cite{zpli95}, a framework  based on the chiral
quark model to study the meson photoproductions is developed.
It starts from the low energy QCD Lagrangian\cite{MANOHAR} so that the
meson-quark interaction is chiral invariant, and the low energy theorem in
the threshold pion-photoproduction\cite{cgln} is automatically
recovered\cite{zpli94} with a proper treatment of the center of mass
motion\cite{zpli93}.  By treating the pseudoscalar mesons as Goldstone
bosons that  interact directly with quarks inside baryons,
the quark model provides an unified formalism for all s- and u-channel
resonances, and the number of parameters used in the model are
dramatically reduced.  In principle, only one parameter is needed for all
resonances that contribute to the meson productions.    This marks
significant advance from the traditional theory, in which the effective
couplings among the hadrons are used so that each resonance requires one
additional parameter.   It also makes it possible to provide a consistent
calculation of the meson-photoproduction beyond the threshold region.
Perhaps more important, it provides an unified description for the
pseudoscalar meson photoproductions and highlights the dynamic role by
baryon resonances in each processes. In this paper, we extend our
investigation to the $\eta$ photoproduction.

There are many features of $\eta$ mesons that make the
$\eta$ photoproductions unique in the quark model. The $\eta$ meson is
an isospin zero state, thus only the resonances with isospin $1/2$
contribute in the s and u channel.  It is also a charge neutral particle
so that the contact (seagull) term\cite{zpli95}
that plays a dominant role in the charge meson production does not
contribute, thus enhances the role of
resonances.  Moreover, because the mass of the
resonance $S_{11}(1535)$ is just above the $\eta N$ threshold where the
S-wave is dominant, the $\eta$ photoproduction in the threshold region
provides us very important probe to the structure of the resonance
$S_{11}(1535)$.  Thus, there have been
considerable theoretical and experimental interests in studying the $\eta$
 photoproduction.
New experimental data for the $\eta$ photoproduction
in the threshold region from Bates\cite{dytman}, ELSA\cite{price}, and
Mainz\cite{krusch} have been published recently.  In particular, the data
from the Mainz group provide more systematic behaviour of the $\eta$
production in the threshold region, which has better energy and
angular resolutions, thus enable us to study the properties of the
resonance $S_{11}(1535)$ more precisely.  Therefore, it is very interesting
to note that the helicity amplitude $A^p_{\frac 12}$ extracted from the new
Mainz data\cite{krusch} is much closer to the prediction of the quark
model\cite{close,simon}.  On theoretical side,
the theoretical studies of the $\eta$ photoproduction were mostly in
the framework of Breit-Wigner parametrizations\cite{breit} or coupled channel
isobar models\cite{benn}.  The recent investigation by the RPI group\cite{muko}
has made significant progress in this field,
 in which the effective Lagrangian approach is used so that
the properties of the resonance $S_{11}(1535)$ extracted from the data
are more model independent, and the number of parameters is
reduced considerably.

Traditionally, investigations of meson photoproductions in the framework of
the quark model have concentrated  on the transition amplitudes, in
particular the helicity amplitudes for the electromagnetic transitions and
the partial wave amplitudes for the mesonic decays of baryon resonances.
These amplitudes were extracted from meson photoproduction data by
the phenomenological models, thus less model independent.  Instead of
relying on the transition amplitudes from the phenomenological models, the
quark model approach enable us to study the structure of baryon
resonances directly from the photoproduction data.  Thus, the connection
between meson photoproductions and more fundamental theories based on QCD
can be established.   Because the transition amplitudes in the quark
model have very different energy and momentum dependences  from those in
traditional models, it is by no means trivial if
meson photoproductions can be successfully described by the quark model.
This requires that the transition amplitudes in the model have correct
off-shell behaviour, which they are usually evaluated on-shell.
Our early investigation\cite{zpli95}
 in the Kaon photoproductions has shown that the quark model approach
presents a much better framework to understand the reaction mechanism of
meson photoproductions than many traditional hadronic models, and we shall
show that the results in the $\eta$ photoproduction are equally
encouraging.

The paper is organized as follows.  The general formalism in the quark
model for the $\eta$ photoproduction is presented in the section 2.
We have carried out three different calculations in the section 3; the
first assumes the $SU(6)\otimes O(3)$ symmetry for the baryon wavefunctions
so that only one parameter is required to fit the experimental data, the
second includes the possible configuration mixing effects for the
resonances $S_{11}(1535)$ and $S_{11}(1650)$ with three additional
parameters, and the third calculation is concentrated on fitting the recent
Mainz data to extract properties of the resonance $S_{11}(1535)$.
Because the data from Mainz group are significantly different from the
rest, we fit them separately.  In section 4, we discuss the structure of
the resonance $S_{11}(1535)$ from the $\eta$ photoproduction data in the
threshold region, and highlight the problems yet to be resolved.
Finally, the conclusion is given in the section 5.

\subsection*{\bf 2. General Formalism}

There are two major components
in addition to the calculation of the electromagnetic and strong
transitions of the baryon resonances in the quark model approach, which has
been shown to be crucial in deriving the model independent low energy
theorem in the threshold pion-photoproduction\cite{zpli94}.
First, one has to combine the phenomenological quark model with the chiral
symmetry, this is being achieved by the introduction of the chiral QCD
Lagrangian\cite{MANOHAR} so that the meson transition operators are
chiral invariant.  Second, since a baryon is being treated as a three quark
system, the separation of the center of mass motion from the internal
motion is important to recover the low energy theorem in the threshold pion
photoproduction,  this has been
discussed in detail in the Compton scattering $\gamma N\to \gamma
N$\cite{zpli93}.

The differential cross section in the center of
mass frame is
\begin{equation}\label{13}
\frac {d\sigma^{c.m.}}{d\Omega}=\frac {\alpha_e \alpha_{\eta}
(E^i+M_N)(E^f+M_N)}
{16sM_N^2}\frac {|{\bf q}|}{|{\bf k}|} |{\cal M}_{fi}|^2
\end{equation}
where $\alpha_{\eta}$ is the $\eta NN$ coupling constant, $\alpha_e$ is the
electromagnetic coupling, $\sqrt
{s}=E^i+\omega_\gamma=E^f+\omega_{\eta}$ is
 the total energy in the c.m. frame.  Generally,  the $\eta$ transitions
between the resonances and the nucleon can be expressed in terms of the
$\alpha_{\eta}$, and no additional parameter for each resonance is
required.  Therefore, the coupling constant has been removed from the
matrix element ${\cal M}_{fi}$ so that it becomes dimensionless.
The coupling constant
$\alpha_{\eta}$ is treated as a free parameter because of the
theoretical issues, such as the U1 anomaly, and $\eta$-$\eta^\prime$
mixing.

One can write the matrix element ${\cal M}_{fi}$ in terms of the
CGLN\cite{cgln} amplitudes;
\begin{equation}\label{14}
{\cal M}_{fi}={\bf J \cdot \vep}
\end{equation}
where $\vep$ is the polarization vector, and the current $J$ is
written as
\begin{equation}\label{15}
{\bf J}=f_1 \vsig+ if_2 \frac {(\vsig \cdot {\bf q})({\bf k}\times \vsig)}
{|{\bf q}| |{\bf k}|}+f_3\frac {\vsig \cdot {\bf k}}{|{\bf q}||{\bf k}|
}{\bf q}+f_4\frac {\vsig \cdot {\bf q}}{{\bf q}^2}{\bf q}
\end{equation}
in the center mass frame.  The differential cross section in terms of the
CGLN amplitude is\cite{tabakin}
\begin{eqnarray}\label{16}
|{\cal M}_{fi}|^2= & Re &\bigg \{ |f_1|^2+|f_2|^2-2\cos(\theta)
f_2f_1^*\nonumber
\\ & + & \frac
{\sin^2(\theta)}2 \left [
|f_3|^2+|f_4|^2+2f_4f^*_1+2f_3f_2^*+2\cos(\theta)f_4f_3^*\right ]\bigg \},
\end{eqnarray}
where $\theta$ is the angle between the incoming photon momentum ${\bf k}$
and outgoing $\eta$ momentum ${\bf q}$ in the center of mass frame.
The various
polarization observables can also be expressed in terms of CGLN amplitudes,
which can be found in Ref. \cite{tabakin}.

The electromagnetic coupling in the nonrelativistic limit is\cite{zpli95}
\begin{equation}\label{17}
h_e=\sum_j e_j\left [r_j\cdot \vep \left (1-\frac {{\bf p}_j\cdot {\bf
k}}{m_q\omega_\gamma}\right )-\frac 1{2m_q}\vsig_j \cdot (\vep \times {\bf
k})\right ],
\end{equation}
and it has been shown\cite{zpli94} that the operator $h_e$ in Eq.
\ref{17} is sufficient to reproduce the low energy theorem for the
threshold pion-photoproductions\cite{cgln}.  The corresponding $\eta$
transition operator is a pseudovector coupling;
\begin{equation}\label{18}
H^{nr}_{\eta}=\sum_j \vsig_j \cdot \left [{\bf A}+\frac
{2\omega_{\eta}}{m_q} {\bf p}_j\right ],
\end{equation}
where ${\bf A}$ corresponds to the center of mass motion, and depends on
the momenta of the initial and final states, and ${\bf p}_j$ is the
internal momentum for a three quark system.

Because the $\eta$ meson is a charge neutral particle, the Seagull term
that plays an important role in the charge meson production does not
contribute.
Thus, the leading Born term would be the nucleon pole term in the
S and U channels;
\begin{eqnarray}\label{27}
{\cal M}_S= \omega_{\eta} e^{-\frac {{\bf q}^2+{\bf k}^2}{6\alpha^2}}
\left ( \fs+\fn\right )\left (1-\frac {{\bf k}^2}{4P^i\cdot k}\mu_N \right )
 \vsig\cdot\vep \nonumber \\  +   ie^{-\frac {{\bf k}^2+{\bf q}^2}
{6\alpha^2}}\left [ \frac {\omega_{\eta}}2\left (\fs+\fn\right)+1\right ]\frac
{\mu_N}{2P^i\cdot k} \vsig
\cdot {\bf q} \vsig \cdot (\vep \times {\bf k})
\end{eqnarray}
where $P^i\cdot k=\omega_{\gamma}(E^i+\omega_{\gamma})$,
 $\mu_N$ is the magnetic moments of the nucleon, and $\alpha^2$ is the
constant from the harmonic oscillator wavefunctions.
The matrix element for
the U-channel nucleon exchange term is
\begin{eqnarray}\label{28}
{\cal M}_U=-e^{-\frac {{\bf k}^2+{\bf q}^2}{6\alpha^2}}
\frac {\mu_N}{2P^f\cdot k}
\bigg \{ \frac {\omega_{\eta}{\bf k}^2}2 \left ( \fs+\fn\right ) \vsig \cdot
\vep +\nonumber \\ i \left [ \frac {\omega_{\eta}}2 \left ( \fs+\fn\right )
+1\right ] \vsig \cdot
(\vep\times {\bf k}) \vsig \cdot {\bf q}\bigg \},
\end{eqnarray}
where $P^f\cdot k=\omega_\gamma (E^f+|{\bf q}|\cos\theta)$.

The contributions from the t-channel exchange are not included in this
approach. This has been discussed in some detail in the
literature\cite{dolen}; if a
complete set of resonances is introduced in the s and u channels, the
inclusion of the t-channel exchange might lead to a double counting problem.
This may turn out to be an advantage of the quark model approach, since
less free parameters are needed to fit the data.

The first excited resonance that contributes to the $\eta$ productions
is the Roper resonance, $P_{11}(1440)$.  In the $SU(6)$ quark model,
its U-channel contribution is
\begin{eqnarray}\label{288}
{\cal M}_{P_{11}(1440)}=\frac {-M_{P_{11}(1440)}{\bf k}^2
e^{-\frac {{\bf q}^2+ {\bf
k}^2}{6\alpha^2}}}{(P^f\cdot k +\delta M_{P_{11}(1440)}^2/2)216m_q\alpha^2}
\bigg \{ \frac {\omega_{\eta}{\bf k}^2{\bf
q}^2}{\alpha^2}\bigg [\fn \nonumber \\
+\fs\bigg ] \vsig \cdot \vep
-i\left \{\frac {\omega_{\eta}}{\mu_q}-\left [ \frac
{\omega_{\eta}}{E^f+M_N}+1\right ]\frac {{\bf q}^2}{\alpha^2}\right \}
\vsig\cdot (\vep \times {\bf k}) \vsig
\cdot {\bf q} \bigg \},
\end{eqnarray}
where $\delta M_{P_{11}(1440)}^2=M_{P_{11}(1440)}^2-M_N^2$, and its
S-channel contribution will be given later.

For the excited resonance with higher energy, such as the P-wave baryons,
we could treat them as degenerate, since their contributions in the
U-channel are much less
sensitive to the detail structure of their masses than
those in the S-channel.  Therefore, we can write their U-channel
contributions in a compact form;
\begin{equation}\label{377}
{\cal M}_U=\left ( {\cal M}_U^3+{\cal M}_U^2\right ) e^{-\frac {{\bf
k}^2+{\bf q}^2}{6\alpha^2}}.
\end{equation}
The first term  in Eq. \ref{377}
represents the process in which the incoming photon and
outgoing $\eta$ meson are absorbed and emitted by the same quarks, it is
\begin{eqnarray}\label{37}
{\cal M}^3= \frac {1}{2m_q}
\left [i{\bf A}\cdot (\vep\times {\bf k})+\vsig\cdot ({\bf
A}\times (\vep\times {\bf k}))\right ]F(\frac {{\bf k}\cdot {\bf q}}{3
\alpha^2}, P^f\cdot k) \nonumber \\
+\frac 1{3}\left [\frac {\omega_{\eta}\omega_{\gamma}}{m_q}\left (1+\frac
{\omega_{\gamma}}{2m_q}\right )\vsig \cdot \vep+\frac
1{\alpha^2}\vsig\cdot {\bf A}\vep\cdot {\bf q}\right ]
  F(\frac {{\bf k}\cdot {\bf q}}{3\alpha^2}, P^f\cdot k+\delta M^2)
\nonumber \\ +\frac {\omega_{\eta}\omega_\gamma}{9\alpha^2m_q}\vsig\cdot
{\bf k} \vep\cdot {\bf q}
  F(\frac {{\bf k}\cdot {\bf q}}{3
\alpha^2}, P^f\cdot k+2\delta M^2).
\end{eqnarray}
where
\begin{equation}\label{33}
{\bf A}=-\omega_{\eta}\left (\fn +\fs\right ){\bf k}-\left ( \omega_{\eta}\fs
+1\right ){\bf q}.
\end{equation}
The function $F(x, y)$ in Eq. \ref{37} corresponds to the product of the
spatial integral and the propagator for the excited states, it can be
written as
\begin{equation}\label{378}
F(x, y)=\sum_n \frac {M_n}{ n! (y+n\delta M^2)} x^n,
\end{equation}
where $n\delta M^2=(M_n^2-M^2)/2$ represents the mass
difference between the ground state and excited states with the major
quantum number $n$ in the harmonic oscillator basis,
 which will be chosen as
the average mass differences between the ground state and the negative
parity baryons so that $\delta M^2\approx 0.74$ GeV$^2$.
The first term in Eq. \ref{37} corresponds to the correlation
between the magnetic
transition and the c.m. motion of the $\eta$ transition operator, it
contributes to the leading Born terms in the U-channel. The second term
in Eq. \ref{37} is the correlations among the internal and c.m. motions of
the photon and $\eta$ transition operators, this term only contributes to the
transitions between the ground and $n\ge 1$ excited states in the
harmonic oscillator basis.
The third term in Eq. \ref{37} corresponds to
the correlation of the internal motions between the photon and $\eta$
transition operators, which only contributes to the transition between the
ground and $n\ge 2$ excited states.
The second term ${\cal M}^2_U$ in Eq. \ref{377} represents the
process in which the  incoming photon
and outgoing $\eta$ are absorbed and emitted by different quarks,  and we
found that
\begin{equation}\label{38}
{\cal M}^2_U=0.
\end{equation}
This is a direct consequence of the isospin couplings.  Eq. \ref{37} can
be summed up to any quantum number $n$,
however, the excited states with large quantum number $n$ become
less significant for the U-channel resonance contributions.  Thus, we only
include the excited states with $n\le 2$, which is the minimum number
required for the contribution from every term in Eq. \ref{37}.

For the S-channel resonance processes, the operator ${\bf A}$ in Eq.
\ref{18} should be
\begin{equation}\label{41}
{\bf A}=-\left (\omega_{\eta}\fs+1\right ) {\bf q}
\end{equation}
in the c.m. frame.  The calculation of the S-channel resonance
contributions is similar to that of the U-channel resonance contributions.
However, since the operator ${\bf A}$
is only proportional to the final state momentum ${\bf q}$, the partial
wave analysis can be easily carried out for the S-channel resonances.

In general, one can write the S-channel resonance amplitudes as
\begin{equation}\label{42}
{\cal M}_R=\frac {2M_R}
{s-M_R^2}e^{-\frac {{\bf k}^2+{\bf q}^2}{6\alpha^2}}{\cal O}_R,
\end{equation}
where $\sqrt {s}=E^i+\omega_{\gamma}=E^f+\omega_{\eta}$
is the total energy of the system, and ${\cal O}_R$ is determined
by the structure of each resonance.  Eq. \ref{42} shows that there should be
a form factor,  $e^{-\frac {{\bf k}^2+{\bf
q}^2}{6\alpha^2}}$ in the harmonic oscillator basis, even in the real
photon limit.    If the mass of a resonance is above the
threshold, the mass $M_R$ in Eq. \ref{42}
should be changed to
\begin{equation}\label{43}
M_R^2 \to M_R(M_R-i\Gamma({\bf q})).
\end{equation}
$\Gamma({\bf q})$ in Eq. \ref{43} is the total width of the resonance,
and a function of the final state momentum ${\bf q}$.  For a resonance
decay to a two body final state with orbital angular momentum $l$,
the decay width $\Gamma({\bf q})$ can be written as
\begin{equation}\label{44}
\Gamma({\bf q})= \Gamma_R \frac {\sqrt {s}}{M_R} \sum_{i} x_i
\left (\frac {|{\bf q}_i|}{|{\bf q}^R_i|}\right )^{2l+1}
\frac {D_l({\bf q}_i)}{D_l({\bf q}^R_i)},
\end{equation}
with
\begin{equation}\label{441}
|{\bf q}^R_i|=\sqrt{\frac
{(M_R^2-M_N^2+M_i^2)^2}{4M_R^2}-M_i^2},
\end{equation}
and
\begin{equation}\label{442}
|{\bf q}_i|=\sqrt{\frac
{(s-M_N^2+M_i^2)^2}{4s}-M_i^2},
\end{equation}
where $x_i$ is the branching ratio of the resonance decaying into a
meson with mass $M_i$ and a nucleon, and $\Gamma_R$ is the total decay width
of the S-channel resonance with the mass $M_R$.  The
function $D_l({\bf q})$ in Eq. \ref{44}
is called fission barrier\cite{bw}, and wavefunction
dependent; here we use
\begin{equation}\label{45}
D_l({\bf q})=exp\left (-\frac {{\bf q}^2}{3\alpha^2}\right ),
\end{equation}
which is independent of $l$.
A similar formula used in I=1 $\pi\pi$ and and p-wave $I=1/2$
$K\pi$ scattering was found in excellent
agreement with data in the $\rho$ and $K^*$ meson region\cite{barnes}.
Generally, the  resonance decays are dominated by the pion channels,
except the resonance $S_{11}(1535)$ whose branching ratio of $\eta N$
channel is around 50 percent. Therefore, we simply set
$x_{\pi}=x_{\eta}=0.5$ for the resonance $S_{11}(1535)$, while
$x_{\pi}=1.0$ for the rest of the resonances as a first order
approximation.

The operator ${\cal O}_R$ in Eq. \ref{42} can be generally written as
\begin{equation}\label{46}
{\cal O}_R=A\left [f_1^R \vsig\cdot \vep
+ if_2^R {(\vsig \cdot {\bf q})\vsig \cdot ({\bf k}\times \vep)}
+f_3^R{\vsig \cdot {\bf k}}\vep\cdot {\bf q}+f_4^R{\vsig \cdot {\bf q}}
\vep\cdot {\bf q}\right ]
\end{equation}
for the pseudoscalar meson photoproduction, where A is the meson decay
amplitude and $f_i^R$ ($i=1\dots 4$) is the photon
transition amplitude.
The meson decay amplitude $A$ is
determined by the spatial wavefunction of resonances and the relative
angular momentum between the final decay products.   In Table 1,
we present the amplitude $A$ in the simple harmonic oscillator basis, in
which the amplitude $A$ depends on the total excitation $n$ and and the
orbital angular momentum $L$.   The  relative angular momentum of the
final decay products is expressed in terms of
the partial wave language in Table 1, in which the $S$, $P$, $D$ and $F$ waves
denote the relative angular momentum 0, 1, 2 and 3 between the
final decay products.   The decay amplitude $A$ in Table 1
is the same as the expression in Table 1 in Ref. \cite{simon}
with $g-\frac 13h=\frac
{|{\bf A}|}{|{\bf q}|}$, and $h=\frac {\omega_{\eta}}{m_q}$.  Note that
${\bf A}$ has a negative sign,  this is consistent with the fitted
value for $g-\frac 13h$ and $h$ in Ref. \cite{simon}.

The photon transition amplitudes $f_i^R$ in Eq. \ref{46} are written in
terms of the CGLN amplitudes, which are shown in Table 2.
They are usually expressed in terms of helicity amplitudes, $A_{1/2}$
and $A_{3/2}$, and the connection between the two representations can be
established. A very important example is the vanishing helicity amplitudes
for the transitions between the resonances belonging to $(70, ^4N)$
representation and protons
due to the Moorhouse selection rule\cite{moor} if one uses the
nonrelativistic transition operator in Eq. \ref{17},
 consequently, the CGLN amplitudes for these resonances are zero as well.
There are 3 important negative parity baryons that belong to $(70, ^4N)$
multiplet in the naive quark model; $S_{11}(1650)$, $D_{13}(1700)$ and
$D_{15}(1675)$. However, it has been shown in the potential quark model
calculation\cite{IK79} that the two states, ${\bf 70}N(^2P_M){\frac
12}^-$ and ${\bf 70}N(^4P_M){\frac 12}^-$, are strongly mixed.
Therefore, the contribution from the resonance $S_{11}(1650)$ to the $\eta$
photoproduction will be studied by fitting to the experimental data.
Indeed, this will provide us a direct insights into the configuration mixing
in the potential quark model.

The CGLN amplitudes for the resonances with total spin 1/2 can be easily
related to the helicity amplitude $A_{1/2}$ that has been frequently
calculated in the quark model.  Only the CGLN amplitude $f_1^R$ is nonzero
for the resonance $S_{11}(1535)$, and this corresponds to $E^+_0$
multipole transition\cite{tabakin}.  Moreover, the amplitudes $f_1^R$
for the S-wave resonances have the same structure as the corresponding
helicity amplitude $A_{1/2}$ in Ref. \cite{zpli90},
in which the same nonrelativistic  transition operator is used.
For the $P$ wave resonances, such as the resonances $P_{11}(1440)$ and
$P_{11}(1710)$, only the CGLN amplitude $f_2^R$ is present,  which gives
a $M^-_1$ transition.  Notice that the resonances with isospin 3/2 do not
contribute to the $\eta$ photoproduction due to the isospin coupling
between the $\eta$ meson and the nucleon.  These results provide an
important consistency check for the CGLN
amplitudes in Table 2.

If one intends to calculate the reaction beyond 2 GeV in the center of mass
frame, the higher resonances with quantum number $n=3$ and $n=4$ must be
included.  Instead, we adopt an approach that treats the resonances
for $n \ge 3$ as degenerate, the sum of the transition
amplitudes from these resonances can be obtained through the approach in
Ref. \cite{zpli93}.  The transition amplitude for the nth harmonic
oscillator shell is
\begin{equation}\label{48}
{\cal O}_{n}={\cal O}_n^2 +{\cal O}_n^3
\end{equation}
where the amplitudes ${\cal O}_n^2$ and ${\cal O}_n^3$ have the same
meaning as the amplitudes ${\cal M}_U^2$ and ${\cal M}_U^3$ in Eqs.
\ref{37} and \ref{38}, and we have
\begin{eqnarray}\label{49}
{\cal O}^3_n
= -\frac {1}{2m_q}
\left [i{\bf A}\cdot (\vep\times {\bf k})-\vsig\cdot ({\bf
A}\times (\vep\times {\bf k}))\right ]\frac 1{n!}\left (\frac
{{\bf k}\cdot {\bf q}}{3\alpha^2}\right )^n \nonumber \\
+\frac 1{3}\left [\frac {\omega_{\eta}\omega_{\gamma}}{m_q}\left (1+\frac
{\omega_{\gamma}}{2m_q}\right )\vsig \cdot \vep+\frac
1{\alpha^2}\vsig\cdot {\bf A}\vep\cdot {\bf q}\right ]\frac 1{(n-1)!}
  \left (\frac {{\bf k}\cdot {\bf q}}{3\alpha^2}\right )^{n-1}
\nonumber \\ +\frac {\omega_{\eta}\omega_\gamma}{9\alpha^2m_q}\vsig
\cdot {\bf k}\vep\cdot {\bf q}
\frac 1{(n-2)!}
  \left (\frac {{\bf k}\cdot {\bf q}}{3
\alpha^2}\right )^{n-2}
\end{eqnarray}
and
\begin{equation}\label{63}
{\cal O}^2_n=0.
\end{equation}
Generally, the resonances with large quantum number $n$ become important as
the energy increases.  Furthermore,  the higher partial wave resonances
with orbital angular momentum $L=n$ become dominant, which correspond to
the resonance $G_{17}(2190)$ for $n=3$ and $H_{19}(2250)$ for $n=4$ for the
$\eta$ photoproductions.  Indeed, only these higher partial wave resonances
can be seen experimentally, and this is consistent with the quark model
predictions. Thus, we simply take the masses and decay
widths of these high partial wave resonances as input in Eq. \ref{42}.

\subsection*{\bf 3. The Numerical evaluation}
We shall take the same procedure as that in the calculation of the kaon
photoproduction\cite{zpli95}.  In order to take into account of the
relativistic effects, the Lorentz boost factor is introduced in the CGLN
amplitudes
\begin{equation}\label{64}
f_i({\bf k},{\bf q}) \to \frac {M_N^2}{E_N^iE_N^f}f_i (\frac {M_N}{E^i_N}{\bf
k}, \frac {M_N}{E_N^f} {\bf q})
\end{equation}
where $i=1\dots 4$, and $\frac {M_N}{E_N^i}$ ($\frac {M_N}{E_N^f}$)
is a Lorentz boost factor for the initial (final) state.

The parameters in this calculation have standard values in the quark model,
which the quark mass $m_q$ is $0.34$ GeV, and $\alpha^2=0.16$ GeV$^2$.
The masses and the decay widths for the S-channel resonances
are taken from the recent particle data group\cite{pdg94}.
In principle, there is only one parameter $\alpha_{\eta}$ to be determined
in the numerical evaluation, in which the wavefunctions of the resonances
are assumed to have $SU(6)\otimes O(3)$ symmetry.  However, one should not
expect that quark model in the $SU(6)\otimes O(3)$ symmetry limit
 could provide a
quantitative description of the $\eta$ production, since there should be
significant configuration mixing\cite{IK79}. In particular, the
configuration mixing between the states $N(^2P_M){\frac 12}^{-1}$ and
$N(^4P_M){\frac 12}^{-1}$ generates a nonzero contribution from the
resonance $S_{11}(1650)$, thus affects the $\eta$ photoproduction in the
threshold region significantly. The evaluations in the potential quark
model\cite{IK79} show that this mixing is indeed very strong.
Therefore, we introduce two parameters, $C_{S_{11}(1535)}$ and
$C_{S_{11}(1650)}$, to take into account of the configuration mixing effects.
The contributions from the resonances $S_{11}(1535)$ and $S_{11}(1650)$
become
\begin{equation}\label{65}
{\cal O}_{S_{11}(1535)}=C_{S_{11}(1535)}{\cal O}_{N(^2P_M){\frac 12}^{-1}},
\end{equation}
and
\begin{equation}\label{66}
{\cal O}_{S_{11}(1650)}=C_{S_{11}(1650)}{\cal O}_{N(^2P_M){\frac 12}^{-1}},
\end{equation}
where ${\cal O}_{N(^2P_M)}{\frac 12}^{-1}$ is given in Table 2.
Furthermore, the U-channel contributions given in Eq. \ref{377} represent
the result in the $SU(6)$ symmetry limit, which correspond to
$C_{S_{11}(1535)}=1$ and $C_{S_{11}(1650)}=0$.  There should be
an additional U-channel contribution for the  general coefficients
$C_{S_{11}}$, and it is given by
\begin{eqnarray}\label{67}
{\cal M}_{S_{11}}^U=\frac {-M_{S_{11}}\omega_{\gamma}
e^{-\frac {{\bf q}^2+ {\bf
k}^2}{6\alpha^2}}}{P^f_N\cdot k +\delta M_{S_{11}}^2/2}\frac
{C_{S_{11}(1535)}+C_{S_{11}(1650)}-1}{3}
\bigg \{ \frac {{\bf q}^2}{3\alpha^2}\left
[ \frac {\omega_{\eta}}{E^f_N+M_N}+1\right ] \nonumber \\
-\frac {\omega_{\eta}}{m_q} +\frac {{\bf q}\cdot {\bf
k}}{3\alpha^2}
\left [\frac {\omega_{\eta}}{E^f+M_N}+\frac {\omega_{\eta}}{E^i+M_N}
\right ]  \bigg \} \left (1+\frac {\omega_{\gamma}}{2m_q}\right )\vsig
\cdot \vep ,
\end{eqnarray}
where $M_{S_{11}(1535)}\approx M_{S_{11}(1650)}\approx 1.6$ GeV in the
U-channel.   In principle, the coefficients $C_{S_{11}}$ could be obtained from
the potential quark models that reproduce the baryon spectroscopy.  On the
other hand, the coefficients $C_{S_{11}}$ obtained from the fitting procedure
could provide an important test to the wavefunctions in the potential quark
models. Moreover, whether the quark model could reproduce the large
branching ratio for the resonance $S_{11}(1535)$ decaying into $\eta N$
channel is still an open question.  Therefore, we adopt two
approaches in the numerical evaluation.  First,  the wavefunctions of the
resonances are assumed to be in exact symmetry limit, thus only one parameter
$\alpha_{\eta}$ is needed to fit the data.  Second, we treat
the coefficients, $C_{S{11}(1535)}$ and $C_{S_{11}(1650)}$, and the total
decay width of the resonance $S_{11}(1535)$, $\Gamma_{S_{11}(1535)}$
 as free parameters, and fit them to the differential cross section
data.

The function minimization routine\cite{numre} is used to minimize the
least square function
\begin{equation}\label{68}
\chi^2=\sum_i\frac {[X_i-Y_i(a_1,\dots, a_n)]^2}{\sigma_{X_i}^2},
\end{equation}
where $X_i$ represents the experimental data, $\sigma_{X_i}$ corresponds to
the error of the data, and $Y_i(a_1,\dots,a_n)$ is the theoretical
predictions with parameters $a_1,\dots,a_n$ to be fitted.
There are about 150 points of differential cross section data  up to
$E_{lab}=1.45$ GeV from the
old data set\cite{landolt}, and recent data by Homma {\it et
al}\cite{homma} and by Dytman {\it et al}\cite{dytman}.  More recently, the
new experimental data in the threshold region from the Mainz group has been
published\cite{krusch}.  This set of data differs significantly from the
old set data\cite{landolt}, and recent Bates data\cite{dytman} in the
threshold region.
 Therefore, we shall fit the Mainz data separately, and the parameters
obtained in these fits are summarized in Table 3.
Although there are also few target polarization data\cite{heusch},
they will not be used in our fitting because these data have
large errors and do not present any systematic behaviour on the target
polarization.

In Fit 1, we assume that the resonances have exact $SU(6)\otimes O(3)$
symmetry, and the masses and decay widths of resonances come from the
recent particle data group\cite{pdg94}.  Therefore, there is only one
parameters, $\alpha_{\eta}$, left to fit the data, and we find
\begin{equation}\label{69}
\alpha_{\eta}=0.465
\end{equation}
by fitting it to the combinations of old data\cite{landolt,homma}
 and recent data from Bates\cite{dytman}.
We present the energy dependence of the differential cross sections
 at $\theta_{cm}=50^\circ \pm 5^\circ$ in Fig. 1, at
$\theta_{cm}=90^\circ \pm 8^\circ$ in Fig. 2, and the energy dependence of the
total cross section is shown in the Fig. 3.  Considering that there is only
one parameter in the calculation, the overall
agreement with the data is truly remarkable.   Our calculation in the
symmetry limit also shows that the resonance $S_{11}(1535)$ is less dominant
than the data suggest, and the calculated total cross section is
significantly larger than the data in $E_{lab} \approx 0.9 - 1$ GeV
region.  This suggests that the resonance $S_{11}(1650)$ also plays a
significant role in addition to the dominant presence of the resonance
$S_{11}(1535)$.  Therefore, we treat the coupling constant $\alpha_{\eta}$,
two coefficients $C_{S_{11}}$, and the decay width $\Gamma_{S_{11}(1535)}$
as a free parameter in Fit 2, the resulting parameters are shown in Table
3. The resulting fits are also shown in Figs. 1, 2, and 3 respectively.
 It is worth to mention that the smaller total decay width
$\Gamma_{S_{11}(1535)}=0.111$ GeV is largely due to
the recent Bates data\cite{dytman} at
$E_{lab}=0.729$ GeV.  Since only the differential cross
section data are used in our fits, the recent total cross section
data\cite{price} from ELSA are not used here.  Clearly,  the
threshold region is quite crucial in determining the mass and width of the
resonance $S_{11}(1535)$.  The resonance $S_{11}(1535)$ becomes more
dominant in this fit, and we find a small but negative contribution from
the resonance $S_{11}(1650)$.  This is qualitatively in agreement with the
recent calculation by the RPI group\cite{muko}, in which the effective
Lagrangian approach is used.    On the other hand, the coupling constant
$\alpha_{\eta}$ is significantly reduced from $0.465$ to about $0.14$,
this shows how strongly dependent of the coupling constant $\alpha_{\eta}$
on the behaviour of the resonances $S_{11}(1535)$ and the resonance
$S_{11}(1650)$.  The physical reason behind the large reduction of the
coupling constant $\alpha_{\eta}$ is that the threshold region is dominated
by the resonance $S_{11}(1535)$, which accounts nearly 90 percent of the
total cross section, thus, a small variation in the $S_{11}(1535)$ will lead
to a larger change in the contribution from the Born term.  This shows
that the threshold region alone is not a reliable source
to determine the $\eta NN$ coupling constant $\alpha_{\eta}$.

In Fit 3, we are concentrating on the recent published data from the Mainz
group\cite{krusch}, in which more systematic
differential cross section data are presented from $E_{lab}=0.716$ to
$0.788$ GeV.  It should be pointed out that these data are
significantly larger than the previous data in threshold
region\cite{landolt,dytman,homma}, therefore, further experimental
confirmation is needed.  Because this set of data is concentrated on the
region from the threshold to the mass of the resonance $S_{11}(1535)$,
one could not obtain any reliable information on the contribution from the
resonance $S_{11}(1650)$. Thus, we exclude the contribution from the
resonance $S_{11}(1650)$ by setting the parameter
$C_{S_{11}(1650)}=0.0$, which is the same as that in Fit 1.  Thus,
three parameters, $\alpha_{\eta}$,
$C_{S_{11}(1535)}$ and $\Gamma_{S_{11}(1535)}$, are fitted to the data.
The calculated
total cross section and the data are presented in Fig. 4, the agreement with
the data is excellent.  Moreover, the resulting decay width
$\Gamma_{S_{11}(1535)}$ is found to be $0.198$ GeV, and in very good
agreement with the simple Breit-Winger fit\cite{krusch}.  This provides an
important consistency check of the model.   One could also see the possible
contribution from the resonance $S_{11}(1650)$ at $E_{lab}\approx 0.8$ GeV,
where the data suggest that the total cross section starts to decrease.
More systematic data in $E_{lab}=0.8 \sim 1.0$ GeV region are needed in
order to learn more on the structure of both $S_{11}(1535)$ and
$S_{11}(1650)$.

To highlight the importance of the data in the threshold region in
determining the properties of the resonance $S_{11}(1535)$ and coupling
constant $\alpha_{\eta}$, we show the
differential cross section at $E_{lab}=0.729$ GeV in Fig 5, and
$E_{lab}=0.752$ in Fig. 6.  Notice that there is a significant difference
between the data from Bates\cite{dytman} and Mainz\cite{krusch}
 at $E_{lab}=0.729$ GeV, it leads to
the change from $\Gamma_{S_{11}(1535)}=0.111$ GeV in Fit 2 to $0.198$ GeV
in Fit 3. Resolving this difference in the future experiments is crucial
in understanding the structure of the resonance $S_{11}(1535)$.
 In the same time, the parameter $\alpha_{\eta}$ is changed by a
factor of 3.   In fact, the calculation by the RPI
group\cite{muko} has shown that one could obtain a good fit to the data
in the threshold region for a wide range of coupling constant
$\alpha_{\eta}$.  More systematic data beyond the threshold region
are calling for, particularly in the region $E_{lab}=1.15 $ to $1.45$ GeV, in
which there is no resonance dominant so that contribution from
the Born term becomes relatively important.
It is interesting to note that the calculated differential cross section in
Fit 2 is
in good agreement with the Bate data\cite{dytman} at $E_{lab}=0.729$ GeV,
but smaller at $E_{lab}=0.752$ GeV, while the results in Fit 3 give
excellent fits to the Mainz data\cite{krusch} in both cases.

We should also point out that the resonances $P_{11}(1710)$ and
$P_{13}(1720)$ also play quite important role in the region $E_{lab}=0.9
\sim 1.1 $ GeV,  and one should not expect that the quark model in the
symmetry limit would provide a quantitative description of these
resonances. One could also study these resonances by inserting coefficients
in front of their CGLN amplitudes and fitting them to the data.  Our
calculation provides a framework to  study the resonance contributions in
the meson photoproduction with less parameters.  This will be investigated
in the future with more accurate data in this region.

The calculation of the target polarizability has also been done with the
parameters in each fit.  We found consistent small polarizabilities at
$\theta_{cm}=90^\circ$ from the threshold to $E_{lab}=1.0$ GeV.  If the
polarization is indeed large in this region as the data
suggest\cite{heusch}, it might
be the evidence that the t-channel meson exchange is required.

\subsection*{\bf 4. The Structure of the Resonance $S_{11}(1535)$}

There were probably two important motivations to study the $\eta$
photoproduction in the threshold region; to determine the $\eta NN$
coupling constant $\alpha_{\eta}$ and to study the structure of the
resonance $S_{11}(1535)$.  Our calculation shows that the threshold region
might not be a reliable source to determine the coupling constant
$\alpha_{eta}$ because of the dominance of the resonance $S_{11}(1535)$.
However, one might be able to learn more about the structure of the
resonance $S_{11}(1535)$ in the quark model.  The study\cite{muko}
 by the RPI group shows that  one could determine the quantity $\xi$ from
the $\eta$ photoproduction, which is defined as
\begin{equation}\label{70}
\xi=\sqrt{\chi^\prime \Gamma_{\eta}}A_{\frac 12}/\Gamma_{T}
\end{equation}
where $\chi^\prime=M_Nk/qM_R$.  One can obtain an analytic expression of the
quantity $\xi$ from the CGLN amplitude in Table 2, which is given by
\begin{eqnarray}\label{71}
\xi=\sqrt{\frac {\alpha_{\eta}\alpha_e\pi(E^f+M_N)}{M_R^3}}\frac
{C_{S_{11}(1535)}\omega_{\gamma}}{6\Gamma_T}\left [\frac
{2\omega_{\eta}}{m_q}-\frac {2{\bf
q}^2}{3\alpha^2}\left (\frac {\omega_{\eta}}{E^f+M_N}+1\right )\right
] \nonumber \\
\left (1+\frac {|{\bf k}|}{2m_q}\right )e^{-\frac {{\bf q}^2+{\bf
k}^2}{6\alpha^2}}.
\end{eqnarray}
Because ${\bf q}^2\ll 1$, the quantity $\xi$ for the resonance
$S_{11}(1535)$ is not sensitive to the parameter $\alpha^2$ related to the
internal structure of the baryon wavefunctions.
After including the Lorentz boost factors in Eq. \ref{64}, we obtain
\begin{equation}\label{72}
\xi= \left \{ \begin{array}{r@{\quad}l}
0.186 & \mbox{From Fit 1} \\
0.208 & \mbox{From Fit 2} \\
0.220 & \mbox{From Fit 3} \end{array} \right.
\end{equation}
in the unit of GeV$^{-1}$.  This is indeed in good agreement with the result
$\xi=0.22\pm 0.02$ GeV$^{-1}$ in ref. \cite{muko}.  It is not surprising
that the quantity $\xi$ in Fit 1 is smaller, because the resonance
$S_{11}(1535)$ is less dominant in the symmetry limit than the data
suggested.  Therefore, assuming that
the branching ration for the resonance $S_{11}(1535)\to \eta N$
is around 0.5, we have the helicity amplitude;
\begin{equation}\label{73}
A^p_{\frac 12}= \left \{ \begin{array}{r@{\quad}l}
81 & \mbox{From Fit 1} \\
78 & \mbox{From Fit 2} \\
111 & \mbox{From Fit 3} \end{array} \right.
\end{equation}
in the unit of $10^{-3}$ GeV$^{-\frac 12}$,
which are very consistent with the results of Ref. \cite{muko} in Fit 1 and
2, and of Ref. \cite{krusch} in Fit 3.

The advantage of the quark model calculation is that the helicity amplitude
$A^p_{\frac 12}$ and the decay width $\Gamma_{S_{11}(1535)}$ can be
predicted separately.  In the symmetry limit, the helicity amplitude
$A^p_{\frac 12}$ is given by\cite{zpli90}
\begin{equation}\label{74}
A^p_{\frac 12}=\sqrt{2\alpha_e\pi\omega_{\gamma}}\frac 1{3\alpha}\left
(1+\frac {|{\bf k}|}{2m_q}\right )e^{-\frac {{\bf k}^2}{6\alpha^2}},
\end{equation}
and the decay width in the $S_{11}(1535)\to \eta N$ channel is expressed in
terms of the coupling constant $\alpha_{\eta}$;
\begin{equation}\label{75}
\Gamma_{S_{11}(1535)}(\eta N)=\frac {\alpha_{\eta}(E+M_N)|{\bf
q}|}{2M_R}\frac {\alpha^2}{M_N^2}\left \{\frac {\omega_{\eta}}{m_q}-\frac
{{\bf q}^2}{3\alpha^2}\left [\frac {\omega_{\eta}}{E+M_N}+1\right ]\right
\}^2e^{-\frac {{\bf q}^2}{3\alpha^2}}.
\end{equation}
We have
\begin{equation}\label{76}
A^p_{\frac 12}=148 \mbox{ $10^{-3}$ GeV$^{-\frac 12}$},
\end{equation}
and
\begin{equation}\label{77}
\Gamma_{S_{11}(1535)}(\eta N)=23.4 \mbox{ MeV}
\end{equation}
after including the Lorentz boost factors.  Comparing this predictions with
the results in Fit 1,  the helicity
amplitude $A^p_{\frac 12}$ predicted by the quark is twice larger than
the data, while the decay width is about factor 3 smaller.
The fact that there is a factor 2 between the old data and the quark model
calculations for the helicity amplitude $A^{p}_{\frac 12}$ has been known
for some time\cite{simon,close}, and it was speculated that this might be
an indication of the configuration mixing\cite{fh82}.  However, the
systematic calculations with the configuration mixings in the Isgur-Karl
model shown\cite{zpli90} that the configuration mixing effects are unable
to reduce the helicity amplitude $A^p_{\frac 12}$.  Therefore, it is
particularly interesting that the new data set from Mainz has
bring the helicity amplitude $A^p_{\frac 12}$ in Fit 3
much closer to the quark
model predictions; the simple Breit-Wigner fit in Ref. \cite{krusch}
also gives
\begin{equation}\label{78}
A^p_{\frac 12}=(125\pm 25)\quad 10^{-3} \mbox{GeV}^{-\frac 12},
\end{equation}
which is even closer to the quark model result in Eq. \ref{76}.

On the other hand, the large $\eta N$ branching ratio for the resonance
$S_{11}(1535)$  has not been fully understood.
The calculation by Koniuk and Isgur\cite{simon} also shows that
the transition amplitude for $S_{11}(1535)\to \eta N$ is about 50 percent
smaller than the data,  whose calculation is similar to this approach.
This is consistent with the fitted coefficient $C_{S_{11}(1535)}\approx
1.5$ in both Fit 2 and 3, and it is unlikely that this can be explained by
the configuration mixing effects.   One of the effects that has not been
taken into account in this study is the finite size of $\eta$ mesons,
which is the one of the major motivations of the quark pair creation
model\cite{LY}.  The calculation has been partly done in Ref. \cite{simon1},
which gives a much larger decay amplitude.
The problem is that the $\eta NN$ coupling that can also be obtained in
this approach was not given, thus there is no base to judge if the parameter
used in the calculation is reasonable.

The understanding of the $\eta N$ branching ratio of the
resonance $S_{11}(1535)$ may be the key to its underlying structure.
It has been discussed  for some time in the literature that the resonance
$\Lambda(1409)$ might be a Kaon-nucleon binding  state, whose mass is just
below the kaon-nucleon threshold.
Notice that the mass of the resonance $S_{11}(1535)$ is just below the
threshold of Kaon
production, $\gamma N\to K\Lambda$ and $\gamma N\to K\Sigma$,  it would be
interesting to study the possibility that the resonance $S_{11}(1535)$ is a
combination of $q^3$ and $K\Lambda$ or $K\Sigma$ binding states, which
suggests that the state $\Lambda (1409)$ as a $K N$ binding state might not
be an isolated case.  The threshold behaviour of Kaon photoproduction,
$\gamma N\to K\Lambda$ and $\gamma N\to K\Sigma$, may provide us further
information in this regard, because it is dominated by the S wave
resonances and the Born terms.

\subsection*{\bf 5. The Conclusion}

The  first quark model calculation is presented for the $\eta$
photoproduction, which provides a very good description of the $\eta$
photoproductions with less parameters.  We show that the threshold region
is not a reliable place to determine the $\eta NN$ coupling constant, which
strongly depends on the properties of the resonance $S_{11}(1535)$.
One should extend the study to $E_{lab}=1.4$ GeV region, in which no
resonance is dominant, so that the contribution from the Born term could be
determined more reliably.
If there is any indication from the recent Mainz data, it might be that the
old set of the data may become irrelevant.  Certainly, the future
experiments planned at various facilities, in particular at CEBAF, will
provide us much more accurate information on the $\eta$ photoproduction
that will reach beyond the threshold region.  The results here show that
the quark model approach will certainly be a very effective tool
of studying the underlying structure of baryon resonances from the
photoproduction data.

\subsection*{Acknowledgment}
Author would like to thank S. Dytman and B. Krusche for providing their
recent experimental information.  Discussions with L. Kisslinger, N.
Mukhopadhyay, R. Shoemaker, F. Tabakin are gratefully acknowledged.
This work is supported by U.S. National Science Foundation grant
PHY-9023586.

\newpage

\begin{tabular}{lcl}
\multicolumn{3}{l}
{Table 1; Meson transition amplitudes $A$ in the simple}\\
\multicolumn{3}{l}{harmonic oscillator basis.}\\[1ex]
\hline\hline
$(N,L)$ & Partial Waves &  $A$ \\ \hline
$(0,0)$ &  P &  $-\left (\frac {\omega_{\eta}}{E^f+M_N}+1\right )$\\[1ex]
$(1,1)$ &  S &  $\frac {2\omega_{\eta}}{m_q}-\left
(\frac {\omega_{\eta}}{E^f+M_N}+1\right
)\frac {2{\bf q}^2}{3\alpha^2}$\\[1ex]
$(1,1)$ & D & $-\left (\frac {\omega_{\eta}}{E^f+M_N}+1\right )$\\[1ex]
$(2,0)$ & P & $\frac {2\omega_{\eta}}{m_q}-\left
(\frac {\omega_{\eta}}{E^f+M_N}+1\right
)\frac {{\bf q}^2}{\alpha^2}$\\[1ex]
$(2,2)$ & P & $\frac {2\omega_{\eta}}{m_q}-\left
(\frac {\omega_{\eta}}{E^f+M_N}+1\right
)\frac {2{\bf q}^2}{5\alpha^2}$\\[1ex]
$(2,2)$ & F & $-\left (\frac {\omega_{\eta}}{E^f+M_N}+1\right )\frac {{\bf
q}^2}{\alpha^2}$\\[1ex] \hline
\end{tabular}

\vspace{2.5cm}

\begin{table}{ Table 2: The CGLN amplitudes for the S-channel baryons
resonances for the proton target in the $SU(6)\otimes O(3)$ symmetry limit,
where $k=|{\bf k}|$, $q=|{\bf q}|$,
and $x=\frac {{\bf k\cdot q}}{kq}$. The CGLN amplitudes
for the $N(^4P_M)$, $N(^4S_M)$, and $N(^4D_M)$ states are zero due to the
Moorhouse selection rule, see text.}\\[1ex]
\begin{tabular}{cllll}\hline\hline
States & $f_1$ & $f_2$  & $f_3$ & $f_4$ \\[1ex]\hline
$N(^2P_M){\frac 12}^-$ & $ \frac {\omega_{\gamma}}{6}\left (1+\frac k
{2m_q}\right ) $ & 0 & 0 & 0\\[1ex]
$N(^2P_M){\frac 32}^-$ & $-\frac {\omega_{\gamma}}{9}\left (1+\frac k
{2m_q}\right )\frac {q^2}{\alpha^2}$ & $-\frac {kqx}
{6m_q\alpha^2}$  & 0 & $\frac {\omega_{\gamma}}{3\alpha^2}$ \\[1ex]
$N(^2S_s^\prime){\frac 12}^+$ & 0 & $-\frac {k^2}{216m_q\alpha^2}$ &
0 & 0 \\[1ex]
$N(^2D_s){\frac 32}^+$ & $\frac {k^2qx}{36\alpha^2}
\left (1+\frac {k}{2m_q}\right )$ & $\frac {
k^2}{216m_q\alpha^2}$ & $\frac {\omega_{\gamma}}{36\alpha^2}$ & 0 \\[1ex]
$N(^2D_s){\frac 52}^+$ & $-\frac {k^2qx
}{180\alpha^2}\left (1+\frac {k}{2m_q}\right )$
& $-\frac {k^2}{144m_q\alpha^2}\left (x^2-\frac 15
\right )$ & $-\frac {
k}{180\alpha^2}$ & $\frac {k^2x}{36q\alpha^4}$ \\[1ex]
$N(^2S_M){\frac 12}^+$ & 0  &$\frac {k^2}{216m_q\alpha^2}$ & 0 &
0\\[1ex]
$N(^2D_M){\frac 32}^+$ & $\frac {k^2qx}{36\alpha^2}
\left (1+\frac {k}{2m_q}\right )$ & $\frac {
k^2}{216m_q\alpha^2}$ & $\frac {k}{36\alpha^2}$ & 0 \\[1ex]
$N(^2D_M){\frac 52}^+$ & $-\frac {k^2qx
}{180\alpha^2}\left (1+\frac {k}{2m_q}\right )$
& $-\frac {k^2}{144m_q\alpha^2}\left (x^2-\frac 15
\right )$ & $-\frac {
k}{180\alpha^2}$ & $\frac {k^2x}{36q\alpha^4}$ \\[1ex]
\hline
\end{tabular}
\end{table}

\vfill

\newpage

\begin{table}
\begin{tabular}{cclclcl}
\multicolumn{7}{l}{Table 3; The parameters obtained from}\\
\multicolumn{7}{l}{different fits. The decay width $\Gamma_{S_{11}(1535)}$}\\
\multicolumn{7}{l}{has a unit of GeV.}\\[1ex]
\hline\hline
 & & Fit 1 & & Fit 2 & & Fit 3  \\[1ex]\hline
$\alpha_{\eta}$ & & 0.465 & & 0.139 & & 0.435 \\[1ex]
$C_{S_{11}(1535)}$ & & 1.0 & & 1.510 & & 1.608 \\[1ex]
$C_{S_{11}(1650)}$ & & 0.0 & & -0.036 & & 0.0 \\[1ex]
$\Gamma_{S_{11}(1535)}$ & & 0.150 & & 0.111 & & 0.198 \\[1ex]
\hline
\end{tabular}
\end{table}
\vfill

\newpage

\subsection*{Figure Captions}
\begin{enumerate}
\item The energy dependence of the differential cross section at
$\theta_{cm}=50^\circ \pm 5^\circ$.  The solid line represents the result
from Fit 1, and the dash line from Fit 2.  The data come from Refs.
\cite{landolt,homma,dytman}.
\item The same as Fig. 1 at $\theta=90^\circ \pm 8^\circ$.
\item The energy dependence of the total cross section, the solid is the
result from Fit 1, and dash line from Fit 2.
\item The result for the total cross section from Fit 3, in which the
parameters are fitted to the data
from Mainz\cite{krusch}.
\item The differential cross section at $E_{lab}=0.729$ GeV.  The solid line
represents the result from Fit 2, and dash line from Fit 2.  The data come
from Ref. \cite{dytman} (triangle) and Ref. \cite{krusch} (circle).
\item The same as Fig 4 at $E_{lab}=0.752$ GeV.
\end{enumerate}

\begin{thebibliography}{99}
\bibitem{zpli95} Zhenping Li, ``The Kaon Photoproduction of Nucleons In The
Chiral Quark Model", hep-ph/9502218, submitted to Phys. Rev. C.
\bibitem{MANOHAR} A. Manohar and H. Georgi, Nucl. Phys. {\bf B234},
189(1984).
\bibitem{cgln} G. F. Chew, M. L. Goldberger, F. E. Low and Y. Nambu,
Phys. Rev. {\bf 106}, 1345(1957); S. Fubini, G. Furlan and C. Rossetti,
Nuovo Cimento, {\bf 40}, 1171(1965).
\bibitem{zpli94} Zhenping Li, Phys. Rev. {\bf D50}, 5639(1994).
\bibitem{zpli93} Zhenping Li, Phys. Rev. {\bf D48}, 3070(1993).
\bibitem{dytman} S. Dytman {\it et al.}, Phys. Rev. {\bf C51}, 2170(1995).
\bibitem{price} J. Price {it et al.}, Phys Rev. {\bf C51}, R2283(1995).
\bibitem{krusch} B. Krusche {\it et al}, Phys. Rev. Lett. {\bf 74},
3736(1995).
\bibitem{close}F. E. Close and Zhenping Li, Phys. Rev. {\bf D42}, 2194(1990).
\bibitem{simon} R. Koniuk and N. Isgur, Phys. Rev. {\bf D21}, 1888(1980).
\bibitem{breit} H. R. Hicks {\it et al}., Phys. Rev. {\bf D7}, 2614(1973),
and refference therein.
\bibitem{benn} C. Bennhold and H. Tanabe, Phys. Lett. {\bf B243}, 12(1990).
\bibitem{muko} M. Benmerrouche, N. C. Mukhopadhyay, and J. F. Zhang, Phys.
Rev. {\bf D51}, 3237(1995). M. Benmerrouche and N. Mukhopadhyay, Phys. Rev.
Lett. {\bf 67}, 101(1992).
\bibitem{tabakin} C. G. Fasano, F. Tabakin and B. Saghai, Phys. Rev. {\bf
C46}, 2430(1992).
\bibitem{dolen} R. Dolen, D. Horn, and C. Schmid, Phys. Rev. {\bf 166},
1768(1966).
\bibitem{bw} J. M. Blatt and V. F. Weisskopf, Theoretical Nuclear Physics,
(Wiley, New York, 1952), p361.
\bibitem{barnes} Zhenping Li, M. Guidry, T. Barnes and E. S. Swanson,
MIT-ORNL preprint, MIT-CTP-2277/ORNL-CCIP-94-01.
\bibitem{moor} R. G. Moorhouse, Phys. Rev. Lett. {\bf 16}, 772(1966).
\bibitem{IK79} N. Isgur and G. Karl,  Phys. Rev. {\bf D18}, 4187(1978);
 N. Isgur and  G. Karl, Phys. Rev. {\bf D19}, 2194(1979).
\bibitem{pdg94} Paticle Data Group, Phys. Rev. {\bf D50}, 1173(1994).
\bibitem{landolt} H. Genzel, P. Joos, and W. Pfeil, in {\it
Landolt-B\"orstein}, New Series I/8 (Springer, New York, 1973).
\bibitem{homma} S. Homma {\it et al.}, J. Phys. Soc. Jpn, {\bf 57},
1381(1988).
\bibitem{heusch} C. A. Heusch, {\it et al.}, Phys. Rev. Lett. {\bf 25},
1381(1970).
\bibitem{numre} W. H. Press {\it et al.}, ``Numerical Recipes", Cambridge
University Press, (New York, 1986), pp. 289.
\bibitem{fh82} F. Foster and G. Hughes, Z. Phys. C{\bf 14}, 123(1982).
\bibitem{zpli90} Zhenping  Li and F.E. Close,
Phys. Rev. {\bf D42}, 2207(1990).
\bibitem{LY} Le Yaouanc {\it et al}, Hadron Transitions In The
Quark Model, (Gordon and Breach, New York, 1988); Phys. Rev. {\bf D8},
2223(1973); {\bf D9}, 1415(1974).
\bibitem{simon1} S. Capstick and W. Roberts, Phys. Rev. {\bf D49}
4570(1994).
\end{thebibliography}
\end{document}